\documentclass[12pt]{article}
\usepackage{graphicx}

\title{\bf CKM 2018
  \\
   Summary of Working Group 4:
  \\
  Mixing and mixing-related CP violation in the B system
  \\
  $\Delta M$, $\Delta \Gamma$, $\phi_s$, $\phi_1/\beta$, $\phi_2/\alpha$, $\phi_3/\gamma$}

\author{Sevda Esen
  \\
  Nikhef, NL
  \\
  E-mail: sevda.esen@cern.ch
  \\
  \\
  Alexander Lenz
  \\
  IPPP, Durham University, UK
  \\
  E-mail: alexander.lenz@durham.ac.uk}

\begin{document}

\clearpage\maketitle 
\thispagestyle{empty}

\begin{centering}
\subsection*{Abstract}
\end{centering}
%\begin{sciabstract}
We summarise the excellent talks of Working group 4 at the 10th CKM Workshop in Heidelberg
taking place from 17th till 21st September 2018.
% \end{sciabstract}

\newpage

\section{B mixing}
\subsection{Introduction}
\begin{figure*}[htb]
  \begin{center}
\includegraphics[width=0.90 \textwidth]{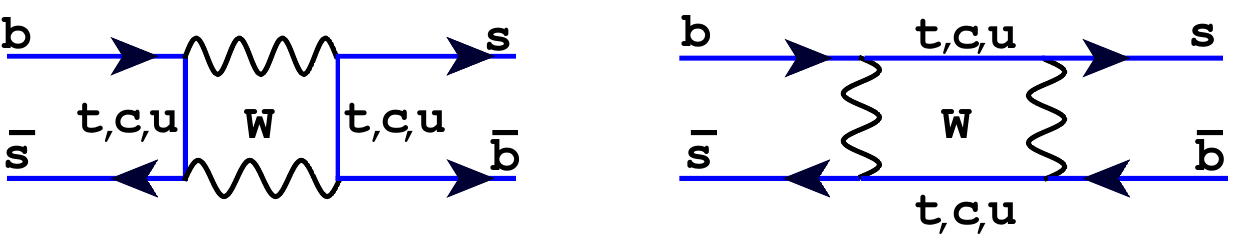}
\caption{\label{box} Standard Model  diagrams for the transition between $B_s$ and $\bar{B}_s$ mesons.}
\end{center}
  \end{figure*}

In the Standard Model (SM) mixing of neutral $B_q$-mesons is governed by the famous box-diagrams,
with internal $W$-bosons
and internal $up$-, $charm$- and $top$-quarks, see Fig. \ref{box} for the case of $B_s$-mesons -
for a more detailed introduction
into $B$-mixing, see e.g. \cite{Artuso:2015swg}.
The contribution of internal on-shell particles
(only the $charm$- and the $up$-quark can contribute) is denoted by
$\Gamma_{12}^q$; the contribution of internal off-shell particles
(all depicted particles can contribute) is denoted by $M_{12}^q$. In the $B$-system there are simple
relations\footnote{This holds not for $D$-mixing, see e.g.
\cite{Jubb:2016mvq,Bobrowski:2010xg,Nierste:2009wg}.}
between $\Gamma_{12}^q$, $M_{12}^q$ and the physical observables mass difference $\Delta M_q$,
the decay rate difference $\Delta \Gamma_q$ and the semi-leptonic asymmetries $a_{sl}^q$:
\begin{equation}
  \Delta M_q  \approx  2 \left|M_{12}^q \right| \, ,
  \hspace{1cm}
  \Delta \Gamma_q \approx  2 \left|\Gamma_{12}^q \right| \cos \phi_{12}^q \, ,
 \hspace{1cm}
  a_{sl}^q  \approx  \left| \frac{\Gamma_{12}^q }{M_{12}^q} \right| \sin \phi_{12}^q \, ,
\end{equation}
with $\phi_{12}^q = \arg (-M_{12}^q/ \Gamma_{12}^q)$.

\subsection{The mass difference $\Delta M_q$}
\begin{center}
\begin{figure*}
  \begin{center}
    \includegraphics[width=0.80 \textwidth]{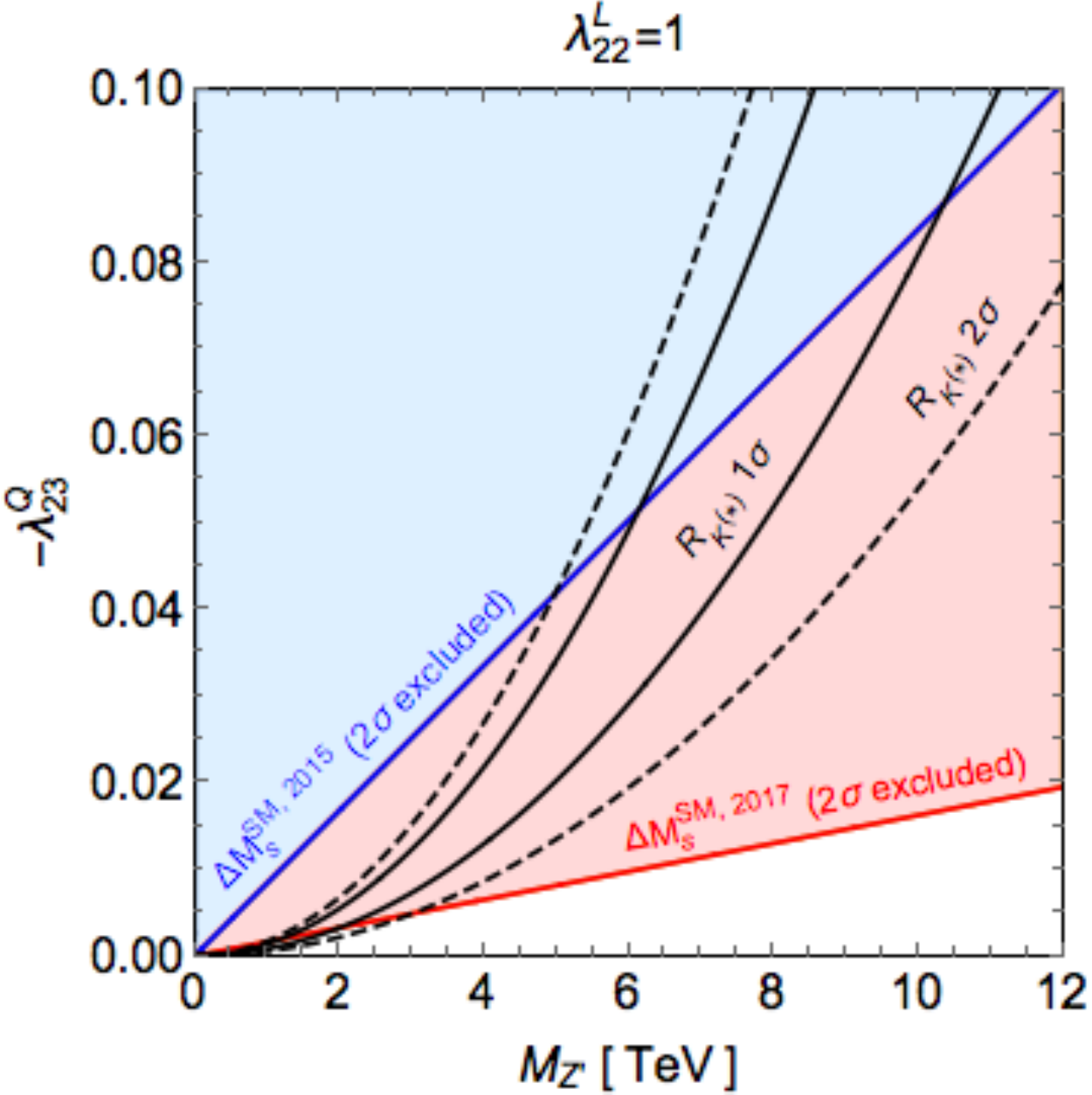}
\caption{\label{bound} Allowed parameter space of $Z'$ models that try to explain the B anomalies.}
  \end{center}
\end{figure*}
\end{center}

\begin{center}
\begin{figure*}
  \begin{center}
    \includegraphics[width=0.70 \textwidth, angle =270]{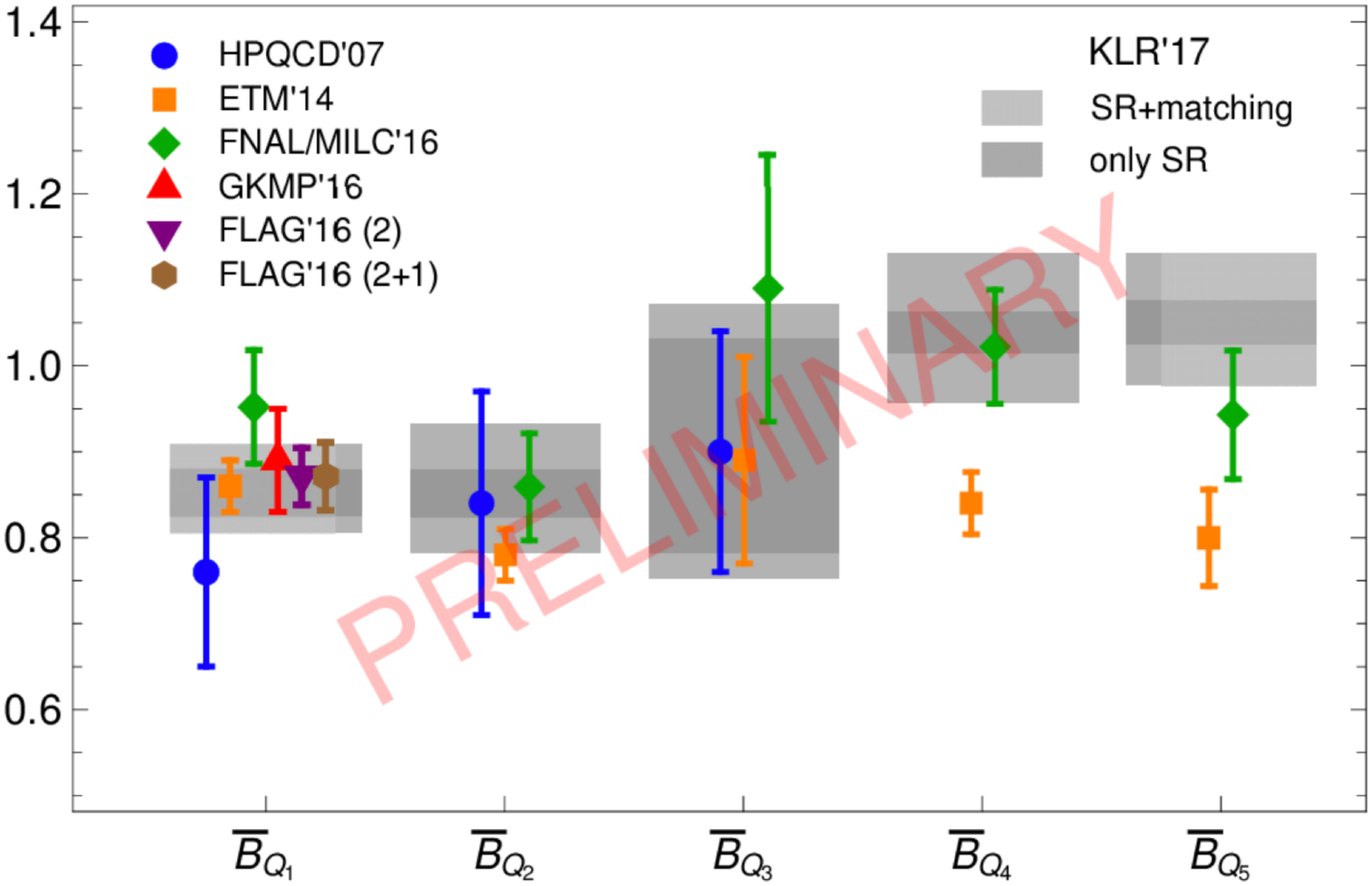}
    \caption{\label{Rauh} Comparison of different determinations of the $B_s$ mixing Bag parameters. The grey areas denote the preliminary values from 
      \cite{ThomasRappear}. The red value (GKMP'16) is the sum rule result from Siegen for $B_d$ mixing \cite{ThomasM}
            and the green symbols denote the FNAL/MILC results.}
  \end{center}
\end{figure*}
\end{center}

Manuel Schiller \cite{Manuel} presented an overview of recent measurements for the mass difference done by LHCb.
These values are known with a high precision \cite{Amhis:2016xyh} (based on the
individual measurements \cite{Abulencia:2006ze,Aaij:2011qx,Aaij:2013mpa,Aaij:2013gja,Aaij:2014zsa}):
\begin{eqnarray}
  \Delta M_d^{\rm Exp.} & = & (0.5065 \pm 0.0019) \,  \mbox{ps}^{-1} \, .
\\
  \Delta M_s^{\rm Exp.} & = & (17.757 \pm 0.021) \,  \mbox{ps}^{-1} \, .
\end{eqnarray}
The  calculation of $M_{12}^q$ gives
\begin{eqnarray}
M_{12}^q & = & \frac{G_F^2}{12 \pi^2} \lambda_t^2 M_W^2 S_0(x_t) \hat{\eta }_B B f_{B_q}^2  M_{B_q} \, ,
\label{M12}
\end{eqnarray}
where $\lambda_t$ denotes the CKM elements $V_{tq}^* V_{tb}$ and the
Inami-Lim function  $S_0$ \cite{Inami:1980fz} contains the result of the 1-loop box diagram in the SM and
perturbative 2-loop QCD corrections \cite{Buras:1990fn} are compressed in the factor $\hat{\eta }_B $.
The bag parameter $B$ and the decay constant $f_{B_q}$
quantify the hadronic contribution to $B$-mixing, the uncertainties of their numerical values make
up the by far biggest uncertainty in the SM prediction of the mass difference. In the
SM only one dimension 6 operator with a V-A Dirac structure arises.
Depending on this input we get a range of predictions for the mass difference in the $B_s$-system as indicated in Table \ref{ME}, taken
from \cite{DiLuzio:2017fdq}.
\begin{center}
  \begin{table*}
    \begin{center}
\begin{tabular}{|c||c|c|}
\hline
$\mbox{Source}$     & $f_{B_s} \sqrt{\hat{B}} $           & $\Delta M_s^{\rm SM} $   
\\
\hline
\hline
HPQCD14 \cite{Dowdall:2014qka}    &  $ (247 \pm 12) \; {\rm MeV} $  & $(16.2 \pm 1.7) \, \mbox{ps}^{-1} $
\\
\hline
HQET-SR \cite{Kirk:2017juj}    &  $ (261 \pm 8) \; {\rm MeV} $  & $(18.1 \pm 1.1)\, \mbox{ps}^{-1} $
\\
\hline
ETMC13 \cite{Carrasco:2013zta}    &  $ (262 \pm 10) \; {\rm MeV} $  & $(18.3 \pm 1.5) \, \mbox{ps}^{-1} $
\\
\hline
HPQCD09 \cite{Gamiz:2009ku} = FLAG13 \cite{Aoki:2013ldr}  &  $ (266 \pm 18) \; {\rm MeV} $  & $(18.9 \pm 2.6)\, \mbox{ps}^{-1} $
\\
\hline
\textbf{FLAG17} \cite{Aoki:2016frl} & \textbf{ $(274 \pm 8) \; {\rm MeV}$} & $(20.01 \pm 1.25)\, \mbox{ps}^{-1}$
\\
\hline
FNAL/MILC 16 \cite{Bazavov:2016nty}  &  $ (274.6 \pm 4) \; {\rm MeV} $  & $(20.1 \pm 0.7) \, \mbox{ps}^{-1} $
\\
\hline
HPQCD06  \cite{Dalgic:2006gp} &  $ (281 \pm 20) \; {\rm MeV} $  & $(21.0 \pm 3.0) \, \mbox{ps}^{-1} $
\\
\hline
RBC/UKQCD14  \cite{Aoki:2014nga} &  $ (290 \pm 20)\; {\rm MeV} $  & $(22.4 \pm 3.4) \, \mbox{ps}^{-1} $
\\
\hline
Fermilab11 \cite{Bouchard:2011xj}  &  $ (291 \pm 18) \; {\rm MeV} $  & $(22.6 \pm 2.8) \, \mbox{ps}^{-1} $
\\
\hline
\end{tabular}
\end{center}
    \caption{List of predictions for the non-perturbative parameter $f_{B_s} \sqrt{\hat{B}}$ and the corresponding SM prediction
      for $\Delta M_s$. The current FLAG average is dominated by the FERMILAB/MILC value from 2016.}
\label{ME}
\end{table*}
\end{center}
Lucia di Luzio \cite{Luca} pointed out the importance of the 
precise value of SM prediction for the mass difference and a strict
control of the corresponding uncertainties. Lepto-quarks and $Z'$
models are popular explanations of the B anomalies\footnote{Due to time and space restrictions we will not attempt
  to cite
  the numerous relevant papers in that field.}; these new models would also affect $B_s$-mixing - in the case of $Z'$ models  already at tree-level.
In Fig. \ref{bound} (from \cite{DiLuzio:2017fdq}) we show the allowed parameter range for a $Z'$ model: in order to explain e.g. $R_{K^{(*)}}$ the mass of the $Z'$
and the coupling to the $b$- and $s$-quark should lie within the black parabola-like shape (the 1 sigma bound is a solid line, the 2 sigma one a dotted line).
Taking the FLAG inputs from 2013 for the mass difference one can exclude the blue region. Taking the new FLAG average, that is dominated by the
2016 FNAL/MILC value we are left with the red exclusion region and almost all of the possible parameter space of the $Z'$ model is excluded. Thus the exact numerical
value of the hadronic input for $B_s$-mixing has severe consequences for beyond the Standard Model (BSM) physics.
Thomas Mannel \cite{ThomasM}, Thomas Rauh \cite{ThomasR} and Aida El-Khadra \cite{Aida} presented in detail three non-perturbative determinations
of the hadronic input for $B$-mixing.
Thomas Mannel described \cite{ThomasM} a series of papers of the Siegen group where perturbative three-loop corrections to a HQET sum rule for the
V-A mixing operator for the $B_d$ mesons were determined \cite{Grozin:2016uqy}.
Later on the matching of the HQET result to QCD was performed at 2-loop \cite{Grozin:2017uto,Grozin:2018wtg}. Interestingly it turns out that
HQET sum rule results are competitive in precision with the most modern lattice determinations of the Bag parameter. Naively one might expect a
precision of about $20 \%$ - $30\%$ for sum rule determinations, but in this case it turns out that one can write down a sum rule for $B-1$
\cite{Grozin:2016uqy,Kirk:2017juj}. Since the value of $B$ is very close to one, a result of e.g. $B-1 \approx -0.10 \pm 0.02...0.03$
transforms in a high precision
for the Bag parameter $B \approx 0.9 \pm 0.02...0.03$.
Unfortunately the mass difference $\Delta M_q$ is proportional to $f_{B_q}^2 B$ and
there is no similar trick known for the decay constant $f_{B_q}$ - so one 
has to use for the sum rule value of the mass difference either a precise determination of the decay constant from lattice simulations or a
much more uncertain value from sum rules \cite{Gelhausen:2013wia}. In Table \ref{ME} the precise lattice value for the decay constant from \cite{Bazavov:2017lyh} was used.
\\
Aida El-Khadra discussed \cite{Aida} in detail the lattice result and the corresponding error budget of the FNAL/MILC collaboration \cite{Bazavov:2016nty},
which dominates the current FLAG average. Their $N_f = 2+1$ determination of $f_{B_s} \sqrt{B}$ achieves an impressive precision of
about $3\%$.
The precision of elder published lattice results is limited by the use of a static action or by smaller ensembles, larger lattice spacings,
fewer configurations,..; but there are plans from RBC/UKQCD (a first paper appeared since the CKM conference \cite{Boyle:2018knm}), HPQCD, ETM,.. to
improve on that and thus cross-check the value of FNAL/MILC, that has severe consequences on BSM effects.
In addition to the SM V-A operator for the mass difference, FNAL/MILC determined also the four remaining $\Delta B =2$ dimension 6 operators of the
so-called SUSY basis. Two of these new operators contribute to the SM prediction of $\Delta \Gamma$, that will be discussed below, all four of the new
operators can appear in BSM contributions to the mass difference.
Two of Aida's main conclusions were:
\begin{itemize}
\item Since the bag parameter $B$ is a derived quantity, it does not profit from correlations. Therefore it is better for phenomenological
      applications to use directly the FLAG averages for the matrix elements, i.e. $f_{B_s} \sqrt{B}$, instead of reconstructing
      them from the averages of bag parameters and decay constants seperatly.
\item There are no roadblocks to increasing the precision of the hadronic $B$-mixing parameter on the lattice.
      With current technology a precision of $1-2\%$ for the dimension 6
      operators seems to be achievable.
\end{itemize}
Thomas Rauh presented \cite{ThomasR} an extension of the sum rule calculation of the Siegen group to all dimension 6 operators for mixing
(see Fig. \ref{Rauh}) and lifetimes of heavy mesons
\cite{Kirk:2017juj} - this will be discussed below - and the determination of $m_s$-corrections \cite{ThomasRappear} to the Siegen result, i.e. the
first sum rule determination of the Bag parameter $B$ for $B_s$ mixing.
Compared to the FNAL/MILC result the new sum rule result for $B_s$ mesons seems to prefer lower values for the Bag parameter, thus being more consistent
with the experimental result for $\Delta M_s$  and leading to weaker constraints for BSM models. Here some words of caution are in order: as Arifa pointed out
FNAL/MILC so far did not perform a direct determination of the Bag parameter - the FNAL/MILC value shown in Fig. \ref{Rauh} is a derived quantity with external
input for the decay constant, the sum rule determination is still preliminary and both approaches still overlap within their uncertainties.
\subsection{The decay rate difference $\Delta \Gamma_s$}
The measurement of the decay rate difference was discussed in the talks of
Manuel Schiller      (LHCb)  \cite{Manuel}, 
Pavel Reznicek       (ATLAS) \cite{Pavel},
Chandiprasad Kar     (CMS)   \cite{Chandiprasad}
and
Varvara Batozskaya  (LHCb) \cite{Varvara}. 
HFLAV combines these measurements to
\begin{eqnarray}
  \Delta \Gamma_s & = & ( 0.088 \pm 0.006) \, \mbox{ps}^{-1} \, ,
  \\
  \frac{\Delta \Gamma_d}{\Gamma_d} & = & -0.002 \pm 0.010 \, .
\end{eqnarray}
The calculation of $\Gamma_{12}^q$ is more involved and is based on the Heavy Quark Expansion (HQE)
(see \cite{Lenz:2014jha} for a review and the original references).
According to the HQE the total decay rate of a
heavy hadron can be expanded in the inverse of the heavy quark mass as
\begin{eqnarray}
  \frac{1}{\tau} = \Gamma & = & \Gamma_0 + \frac{\Lambda^2}{m_b^2}\Gamma_2
  + \frac{\Lambda^3}{m_b^3}\Gamma_3
  + \frac{\Lambda^4}{m_b^4}\Gamma_4 + ... \, .
\end{eqnarray}
The hadronic scale $\Lambda$ is of order $\Lambda^{QCD}$, its numerical value has to be determined
by direct computation.
For hadron lifetimes it turns out that the dominant correction to $\Gamma_0$ is the third term
$\Gamma_3$. Each of the $\Gamma_i$'s can be split up in a perturbative part and non-perturbative
matrix elements - it can be
formally written as
\begin{eqnarray}
  \Gamma_i & = & \left[ \Gamma_i^{(0)} + \frac{\alpha_S}{4 \pi} \Gamma_i^{(1)}
                                      + \frac{\alpha_S^2}{(4 \pi)^2} \Gamma_i^{(2)} + ... \, ,
    \right] \langle O^{d=i+3} \rangle
\end{eqnarray}
where $\Gamma_i^{(0)}$ denotes the perturbative LO-contribution, $\Gamma_i^{(1)}$ the NLO one and so on;
$\langle O^{d=i+3} \rangle$ is the non-perturbative matrix element of $\Delta B = 0$ operators
of dimension $i+3$. The mixing quantity $\Gamma_{12}^q$ obeys a very similar HQE, but now the operators change the $b$-quantum number by two units, $\Delta B = 2$:
\begin{eqnarray}
 \Gamma_{12} & = & 
  \frac{\Lambda^3}{m_b^3}\Gamma_3
  + \frac{\Lambda^4}{m_b^4}\Gamma_4 + ... \, .
\end{eqnarray}
Uli Nierste \cite{Uli} gave an overview of the theoretical status of the SM prediction for $\Gamma_{12}^s$. In the $\overline{MS}$ scheme he obtains
\begin{equation}
\Delta \Gamma_s = \left( 0.104 \pm 0.008_{scale} \pm 0.007_{B} \pm 0.015_{1/m_b} \right) \, \mbox{ps}^{-1} \, ,
\end{equation}
where the first uncertainty is due to an unphysical renormalisation scale dependence at NLO-QCD ($\Gamma_3^{(1)}$)
\cite{Beneke:1998sy,Beneke:2003az,Ciuchini:2003ww}. In the pole mass scheme this uncertainty is considerably larger \cite{Uli} :
\begin{equation}
\Delta \Gamma_s = \left( 0.091 \pm 0.020_{scale} \pm 0.006_{B} \pm 0.017_{1/m_b} \right) \, \mbox{ps}^{-1} \, .
\end{equation}
The scale uncertainty can be reduced by a NNLO-QCD calculation. The second uncertainty is due to the matrix elements of operators of dimension 6.
Here two additional operators to the one appearing in the mass difference are arising. We have currently a HQET sum rule determination for $B_d$ mesons
\cite{Grozin:2016uqy,Kirk:2017juj} and lattice determinations from 2016 \cite{Bazavov:2016nty} ($N_f=2+1$) and 2013 \cite{Carrasco:2013zta} ($N_f=2$).
For Uli's prediction the values from FNAL/MILC have been used. The third uncertainty stems from higher orders in the HQE.
The dimension 7 perturbative part has been determined already in 1996 by Buchalla and Beneke \cite{Beneke:1996gn} for $B_s$ and in \cite{Dighe:2001gc}
for $B_d$ - the non-perturbative matrix elements have so far only been estimated in vacuum insertion approximation. 
\\
Uli presented a first calculation of a sub-set of all NNLO-QCD corrections ($\Gamma_3^{(2)}$)  \cite{Asatrian:2017qaz}
and promised the full $\alpha_s/m_b$ corrections ($\Gamma_4^{(1)}$)  for CKM 2020,
the full NNLO-QCD corrections ($\Gamma_3^{(2)}$) to $\Delta \Gamma_s$ for CKM 2022
and
the full NNLO-QCD corrections ($\Gamma_3^{(2)}$) to the semi-leptonic asymmetries (here also subleading CKM structures have to be determined) for CKM 2024.
\\
Matthew Wingate \cite{Matthew} presented the current status of the ongoing HPQCD activities \cite{Davies:2017jbi} to perform the first non-perturbative
determination of the matrix elements of the dimension 7 operators. These matrix elements are currently only estimated in vacuum insertion and they
therefore make up the largest contribution to the SM uncertainty of $\Delta \Gamma_s$. Any profound non-perturbative determination of the dimension 7
contribution will considerably reduce the theory uncertainty and have many phenomenological implications. Finally it will also be very interesting to see
how well the vacuum insertion works at dimension 7.
\\
We finally had a talk of Gilberto Tetlalmatzi-Xolocotzi 
\cite{Gilberto}, who was questioning the assumption of having no new 
physics effects acting in tree-level non-leptonic decays, see e.g.
\cite{Jager:2017gal}.
According to his studies new effects of the order of $10 \%$ to the 
tree-level Wilson coefficients $C_1$ and $C_2$ are clearly not ruled out
yet. For some observables these small deviations could have dramatic effects:
\begin{itemize}
    \item The decay rate difference of neutral $B_d$ mesons, $\Delta \Gamma_d$ could be enhanced by up to $+160 \% $ or $-291 \% $ 
    of its SM value
    \cite{Jubb:2016mvq}.
    \begin{eqnarray}
    \Delta \Gamma_d^{\rm SM}  & = &
     (2.99 \pm 0.52) \cdot 10^{-3} \mbox{ps}^{-1} \, . 
    \end{eqnarray}   
    \item The experimental extraction of the CKM angle $\gamma$ 
    could be modified by up to $5^\circ$ compared to the SM expectation
    - a huge value compared to the future planned uncertainties \cite{Vos:2019nye}.
\end{itemize}
For completeness we also show the SM predictions for the semileptonic CP asymmetries  \cite{Jubb:2016mvq}.
\begin{eqnarray}
   a_{sl}^s = (2.27 \pm 0.25) \cdot 10^{-5} \, ,
   &&
   a_{sl}^d = -(4.90 \pm 0.54) \cdot 10^{-4} \, .
 \end{eqnarray}

\section{Lifetimes}
\begin{figure*}[htbp]
  \begin{center}
\includegraphics[width=0.90 \textwidth]{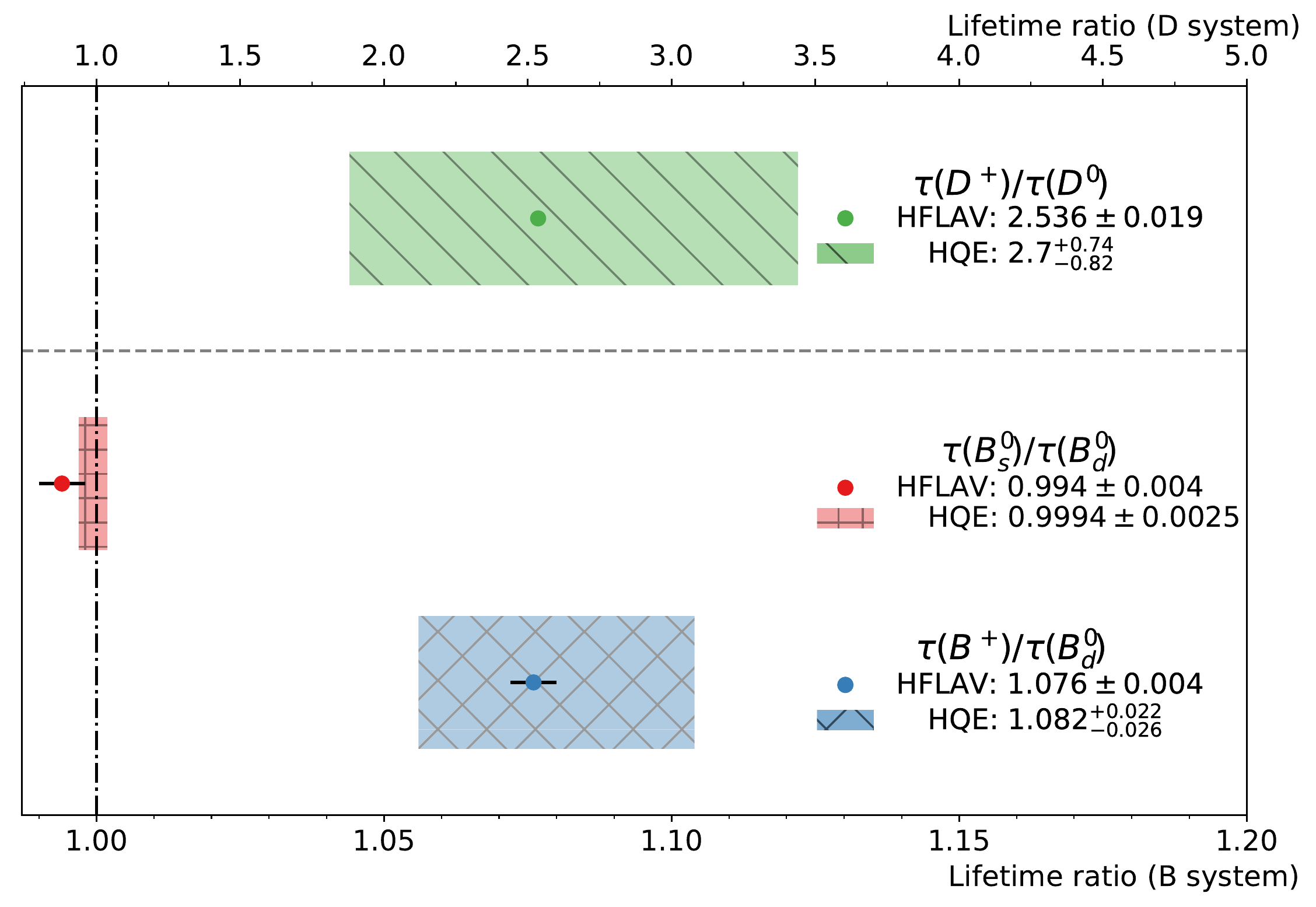}
\caption{\label{lifetime} Comparison of the most solid SM predictions for heavy lifetimes with experiment.}
  \end{center}
\end{figure*}

Lifetime measurements were presented  in the talks of
Manuel Schiller      (LHCb)  \cite{Manuel}, 
Pavel Reznicek       (ATLAS) \cite{Pavel},
Chandiprasad Kar     (CMS)   \cite{Chandiprasad}
and
Varvara Batozskaya  (LHCb) \cite{Varvara}. 
HFLAV combines these measurements to \cite{Amhis:2016xyh}
\begin{eqnarray}
  \frac{\tau (B^+)}{\tau (B_d)}  & = & 1.076 \pm 0.004 \, ,
  \\
  \frac{\tau (B_s)}{\tau (B_d)}  & = & 0.993 \pm 0.004 \, ,
  \\
  \frac{\tau (\Lambda_b)}{\tau (B_d)}  & = & 0.967 \pm 0.007 \, ,
\end{eqnarray}
Thomas Rauh \cite{ThomasR} presented the current status of SM 
predictions depicted in Fig. \ref{lifetime} - see also 
\cite{Lenz:2018ick} for a detailed overview of the current 
theory status. The SM values are based on the perturbative
calculations \cite{Uraltsev:1996ta,Neubert:1996we,Beneke:2002rj,Franco:2002fc,Lenz:2013aua}.
The non-perturbative dimension 6 matrix elements for mesons 
(except for small corrections arising in $B_s$ and $D_s$) were
recently calculated
via HQET sum rules \cite{Kirk:2017juj} - here a complementary lattice evaluation would be very important, either for looking for BSM effects
in the very precisely predicted ratio $\tau (B_s)/ \tau(B_d)$
- this could point towards new effects in hadronic tree-level decays \cite{Jager:2017gal} - , or for testing
the convergence of the HQE in the $b$- and in particular in the $charm$-system.
For baryons we do not have a complete first principle determination
of the non-perturbative matrix elements - there are sum rule determinations of the condensate contribution for the $\Lambda_b$
\cite{Colangelo:1996ta} -  we have, however, some estimates \cite{Lenz:2014jha,Cheng:2018rkz} of the size of the matrix elements using spectroscopy
as an input (based on \cite{Rosner:1996fy}).
LO dimension 7 contributions were determined in
\cite{Lenz:2013aua,Cheng:2018rkz,Gabbiani:2003pq}. 
So far there exists no non-perturbative determination
of the matrix elements of dimension 7 operators.
In Fig. \ref{lifetime}, taken from \cite{Kirk:2017juj},  we compare the most solid SM predictions for heavy lifetimes with experiment and find an
excellent agreement, as well as a strong hint for the convergence 
of the HQE for total inclusive rates in the charm system.

\section{Mixing angles}

%The recent measurements of CP violation in mixing were presented by Mar\'ia Vieites D\'iaz (LHCb) \cite{Maria},  Simon Akar  (LHCb)\cite{Simon},  Bilas Pal (Belle) \cite{Bilas}, Varvara Batozskaya (LHCb) \cite{Varvara} and Pavel  \u Rezn\'i\u cek (ATLAS) \cite{Pavel}. 

\begin{figure*}[htb]
  \begin{center}
\includegraphics[width=0.90 \textwidth]{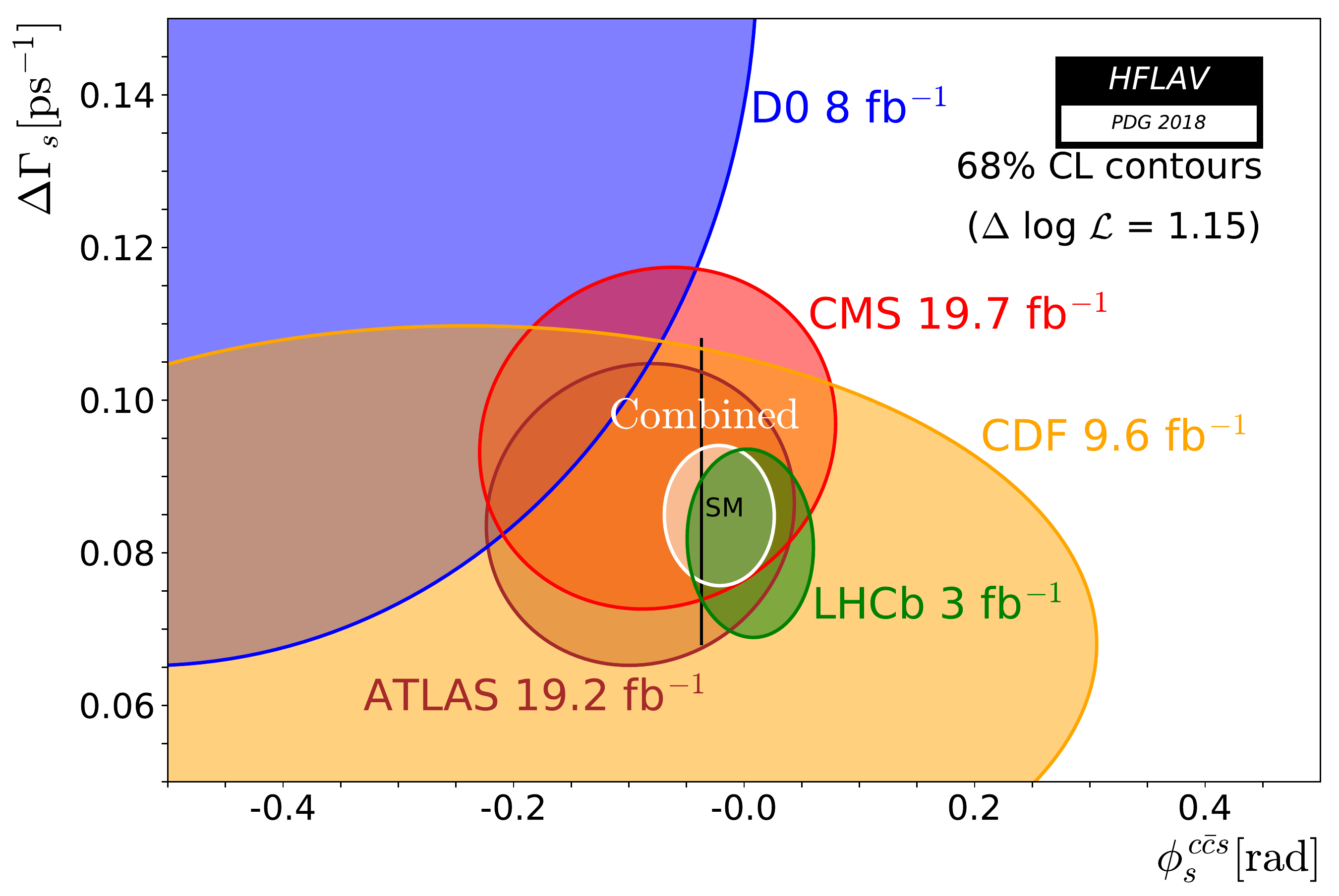}
\caption{\label{phis-DGs} The ($\phi_s^{c\bar{c}s}, \Delta\Gamma_s$) plane, the individual 68\% confidence-level contours of ATLAS, CMS, CDF, D0 and LHCb, their combined contour (white solid line and shaded area), as well as the Standard Model predictions (very thin black rectangle) are shown.}
  \end{center}
\end{figure*}

Neglecting penguin contributions one gets a very precise SM prediction for 
the mixing-induced CP-violating phase $\phi_s^{c\bar{c}s}$ of $-0.0370\pm0.0006$ rad. 
If there is new physics acting in $B$ mixing (i.e. in $M_{12}$), then 
$\phi_s^{c\bar{c}s}$ and $\phi_{12}^d$ (defined below Eq.(1)) receive the
same new contributions - historically this lead regularly to some
confusion
between these two phases \cite{Lenz:2011zz,Lenz:2007nk}.
Recent measurements of $\phi_s^{c\bar{c}s}$ 
by LHCb and ATLAS using Run1 data were presented
in detail by Varvara Batozskaya \cite{Varvara}  and Pavel  \u Rezn\'i\u cek
\cite{Pavel}.  The measurements from LHCb include 
$B_s^0 \to J/\psi K^+K^-,  J/\psi \pi^+\pi^-, \Psi(2S)\phi$ and $ D_s^-D_s^+$.  
The current HFLAV
combination including all presented results reads 
$\phi_s^{c\bar{c}s} = -0.021 \pm 0.031 \, \textrm{rad}$ and
as shown in Fig. \ref{phis-DGs} it is consistent with the SM prediction.
Processes occurring purely via gluonic or electroweak penguin transitions
provide an excellent opportunity to search for new heavy particles
entering in the penguin loops and new results from LHCb for such decays
were presented by Mar\'ia Vieites D\'iaz  \cite{Maria}.   
The effective weak phase $\phi_s^{d\bar{d}s}$  and
$\phi_s^{s\bar{s}s}$ using  flavour-tagged, time-dependent, 
amplitude analyses of  $B_s^0\to (K^+\pi^-)(K^-\pi^+)$ 
\cite{Aaij:2299466} and $B_s^0 \to \phi\phi$ \cite{LHCb:2018iyb}
decays are measured to be
\begin{eqnarray}
 \phi_s^{d\bar{d}s}  &=& -0.10\pm 0.13\pm 0.14 \, \textrm{rad}\,,  
 \\
 \phi_s^{s\bar{s}s} &=& -0.06 \pm 0.05 \pm 0.03\, \textrm{rad}\, .
\end{eqnarray}
The measurement of $\phi_s^{s\bar{s}s}$  is the first CP violation
measurement using run 2 data from LHCb, while $\phi_s^{d\bar{d}s}$ using
only run 1 is the first measurement of this phase. Both results are in
agreement with the SM prediction.  
\\
The CKM angle $\beta(\phi_1)$ is very well measured by B-factories and 
in recent years also by LHCb with a competing precision.  The current
world average is $(22.2 \pm 0.7)^{\circ}$ or $(67.8 \pm 0.7)^{\circ}$.
This two-fold ambiguity has been recently resolved by a joint Belle and
Babar analysis of $B^0 \to D^{(*)0}h^0 (\pi^0, \eta, \omega)$  decays 
and the larger solution is excluded by $7.3\sigma$ \cite{Bilas}.  
\\
The measurements of $\phi_{s,d}(-2\beta)$ from $b\to c\bar{c}s$
transitions involves not only tree-level contributions, but also 
penguin diagrams. These "penguin-pollution" contributions are very
hard to be calculated directly, see \cite{Frings:2015eva}. Therefore
simplifying assumptions like SU(3) flavour symmetries are commonly used
to estimate the potential size of penguin pollution via decay channels
with enhanced penguin-to-tree amplitude ratios. 
The CP violation measurement for the decay $B_s^0\to J/\psi K_S^0$ 
using a time-dependent flavour-tagged analysis with LHCb run 1 data 
set was presented by Simon Akar \cite{Simon}. Combining this result 
with existing  measurements from  $B^+\to J/\psi K^+$,    
$B^+\to J/\psi \pi^+$ and   $B^0\to J/\psi \pi^0$  decays,  
the shift due to the penguin-pollution is estimated to be 
$\Delta\beta = (-1.10^{+0.70}_{-0.85})^\circ$.  
This result will be further improved using updated Belle 
measurements of the branching fraction and  CP asymmetries
of the $B^0\to J/\psi \pi^0$  decay \cite{Bilas}. 
Additionally, a new measurement of $\sin2\beta^{eff}$ 
using $B^0\to K_S\pi^0\pi^0$  decays was presented by 
Bilas Pal and found to be consistent with the measurements 
from $b\to c\bar{c}s$ decays.  
\\
 Benjamin Oberhof presented Belle2 prospects for the mixing 
 and CP violation in B decays \cite{Benjamin}.  With improved 
 detector performance and huge amount of data to be collected 
 by both Belle2 and LHC experiments,  the precision of
 the mixing angles is expected to be better than  
 1\% level by 2030.  
 \\
 Martin Jung presented the theoretical aspects of the precision
 determination of the mixing angles and effects of SU(3) breaking 
 in penguin pollution estimates \cite{Martin} - updating earlier
 results obtained in \cite{Jung:2012mp,Jung:2009pb,Faller:2008zc}.
Martin's conclusion was that $b \to c \bar{c} s$ transitions 
remain golden modes and that $SU(3)_F$ methods will improve with better
data.

\section{Conclusion}
During  CKM 2018 we witnessed an impressive improvement both in experiment
and theory for mixing and mixing related CP observables. A continuation 
of this work will hopefully lead to a deeper understanding of the fundamental principles of  nature.

\section*{Acknowledgement}
We would like to thank all speakers of our session for their excellent talks and the organisers of CKM 2018 for their amazing work.

\end{document}